\documentclass[prc,nofootinbib,preprint,preprintnumbers]{revtex4-1}
\usepackage{epsfig,graphics,color}

\usepackage{bbm}
\usepackage{amssymb}
\usepackage{dcolumn}
\usepackage{amsmath}
\usepackage{mathrsfs}
\usepackage{bbold}
\usepackage{xspace}
\usepackage{slashed}
\usepackage{hyperref}

\usepackage[utf8]{inputenc}

\newcommand{\Nc}{\ensuremath{N_c}\xspace}
\newcommand{\pplus}{\mathbf{p}_+}
\newcommand{\pminus}{\mathbf{p}_-}
\newcommand{\vsig}{\vec{\sigma}}
\newcommand{\vtau}{\vec{\tau}}

\newcommand{\calO}{\mathcal{O}}
\newcommand{\sT}{\slashed{T}}
\newcommand{\sP}{\slashed{P}}
\newcommand{\bg}{\bar{g}}

\begin{document}
\title{Time-reversal-invariance-violating nucleon-nucleon potential in the $1/\Nc$ expansion}

\author{Daris Samart}
\affiliation{Department of Applied Physics,\\
Faculty of Sciences and Liberal Arts,\\
Rajamangala University of Technology Isan,\\
Nakhon Ratchasima, 30000, Thailand}
\affiliation{School of Physics,\\
Suranaree University of Technology,\\
Nakhon Ratchasima, 30000, Thailand }

\author{Carlos Schat}
\affiliation{Departamento de F\'{i}sica, FCEyN,\\
Universidad de Buenos Aires,\\
and IFIBA, CONICET,\\
Ciudad Universitaria,\\
Pabell\'on 1,\\
(1428) Buenos Aires, Argentina
}

\author{Matthias R.~Schindler}
\affiliation{Department of Physics and Astronomy,\\
University of South Carolina,\\
Columbia, SC 29208, USA}

\author{Daniel R.~Phillips}
\affiliation{Institute of Nuclear and Particle Physics and Department of Physics and Astronomy,\\
Ohio University,\\
Athens, OH 45701, USA}

\date{\today}

\begin{abstract}
We apply the large-\Nc expansion to the time-reversal-invariance-violating (TV)
nucleon-nucleon potential.  The operator structures contributing to
next-to-next-to-leading order in the large-\Nc counting are constructed.  For
the TV and parity-violating case we find a single operator structure at
leading order. 
The TV but parity-conserving potential contains
two leading-order terms, which however are suppressed by 1/\Nc compared to the
parity-violating potential. Comparison with phenomenological potentials, including the chiral EFT potential in the TV parity-violating case, leads to large-\Nc
scaling relations for TV meson-nucleon and nucleon-nucleon couplings. 
\end{abstract}

\maketitle


\section{Introduction}
\label{sec:intro}

Time-reversal-invariance violation, or equivalently (assuming the validity of the CPT
theorem \cite{Luders:1954zz,PauliCPT,Bell:1996nh}) CP violation, is an important component in the search for
physics beyond the standard model (BSM).  While the standard model contains
CP-violating mechanisms in the complex phase of the Cabibbo-Kobayashi-Maskawa
matrix and the QCD $\theta$ term, the predicted effects are much smaller than
current experimental bounds on CP-violating observables.  A signal of T
violation beyond these predictions would be a clear indication of BSM physics.
Among the considered time-reversal-invariance-violating (TV) observables, the
neutron electric dipole moment (EDM) has received particular experimental
interest, with the current upper limit $|d_n|< 3.0\times 10^{-26}\, e\,
\text{cm}$ (90\% C.L.) \cite{Baker:2006ts,Afach:2015sja}.  However,  more
information than the measurement of a single observable is necessary to obtain
detailed information about the underlying TV mechanisms.

Additional observables that have been considered include the EDMs of light
nuclei, neutron-nucleus reactions, and nuclear decay parameters (see, e.g.,
Refs.~\cite{Gudkov:1991qg,Pospelov:2005pr,Engel:2013lsa,Bowman:2014fca}).  In all of these
processes, TV nucleon-nucleon (NN) forces play an important role. TV
interactions can either be parity-conserving (PC) or parity-violating (PV),
with the latter expected to give larger contributions to observables such as
EDMs. These forces are the manifestation on the hadronic level of TV interactions among fundamental
degrees of freedom.  Because QCD is nonperturbative at
low energies, a direct derivation of NN forces from the underlying theory is
complicated, so various phenomenological parameterizations of TV NN forces have  been developed. A general
parameterization analogous to Wigner's approach to the T-conserving (TC)
potential \cite{Eisenbud:1941} was given in Ref.~\cite{Herczeg:1966}.  In
phenomenological applications it is common to use a single-meson-exchange
picture, with one strong (TC) and one TV meson-nucleon vertex
\cite{Simonius:1975ve,Gudkov:1992yc,Towner:1994qe,Liu:2004tq}.  More recently,
TV interactions have been constructed in the effective field theory (EFT)
framework, see, e.g.,
Refs.~\cite{deVries:2012ab,Bsaisou:2014oka,Mereghetti:2015rra,deVries:2015gea}
and references therein. In all of these approaches, the TV short-distance
physics is captured in the values of TV coupling constants: either
meson-nucleon couplings and/or NN contact terms.  However, the values of the
couplings have not been derived from the underlying theory and couplings are
constrained only weakly -- if at all -- by experiment.

In the following we apply the 1/\Nc expansion of  QCD
\cite{'tHooft:1973jz,Witten:1979kh} to the TV NN potential, where \Nc is the
number of colors.  The large-\Nc analysis was first applied to the TCPC NN
potential in Refs.~\cite{Kaplan:1995yg,Kaplan:1996rk}, and more recently to
three-nucleon forces \cite{Phillips:2013rsa} and to the TCPV potential
\cite{Phillips:2014kna,Schindler:2015nga}. The large-\Nc expansion analysis allows us to capture 
dominant QCD effects of embedding the fundamental TV interactions in the
nonperturbative environment of the nucleon.  As a result, we find a hierarchy of terms in
the TV potentials: In the  TVPV case there is a single leading-order (LO)
operator structure, with corrections suppressed by a single factor of 1/\Nc.
For TVPC interactions we find two LO terms, with subleading corrections again
suppressed by 1/\Nc.  However, the leading TVPV and TVPC operators do not
contribute at the same order: the dominant TVPV operator contributes at
$\calO(\Nc)$, while the TVPC potential receives contributions starting at
$\calO(1)$.   This hierarchy has to be superimposed on any suppression 
coming from the
underlying BSM physics. At low energies it can be combined with the chiral suppressions 
that originate in the nonperturbative regime of QCD. The large-\Nc and chiral suppressions are
independent and complementary and, given the difficulty in obtaining
experimental constraints, taken together they provide useful additional
theoretical constraints that simplify the analysis of TV observables by
reducing the number of unknowns that need to be considered in phenomenological
applications. 

The paper is organized as follows: Sec.~\ref{sec:largeN} introduces the
framework for analyzing NN potentials in the large-\Nc expansion.  In
Sec.~\ref{sec:potentials} we construct the TVPV and TVPC potentials at leading order (LO),
next-to-leading order (NLO), and next-to-next-to-leading order (NNLO) in the
large-\Nc counting.  These potentials are compared with phenomenological forms
in Sec.~\ref{sec:Compare}, which allows us to extract the large-\Nc scaling of
the various TV couplings.  We conclude in Sec.~\ref{sec:conc}.


\section{The NN potential in the 1/\Nc expansion}
\label{sec:largeN}

Following Ref.~\cite{Kaplan:1996rk}, we define the NN potential as the matrix element
\begin{equation}
\label{eq:PotDef}
V(\pminus,\pplus) = \langle  (\mathbf{p^\prime}_1,C), (\mathbf{p^\prime}_2,D) \vert H  
\vert (\mathbf{p}_1, A) ,(\mathbf{p}_2, B) \rangle.
\end{equation}
Here, $A,\ldots,D$ collectively represent the spin and isospin components of
the nucleons and $\mathbf{p}_i$ ($\mathbf{p^\prime}_i$) denotes the incoming
(outgoing) momentum of the $i$th nucleon, while
\begin{equation}
\mathbf{p}_\pm = \mathbf{p^\prime} \pm \mathbf{p}\,,
\end{equation}
where
\begin{equation}
\mathbf{p} = \frac12\,\left(\mathbf{p}_1 - \mathbf{p}_2\right)\,,\qquad\quad \mathbf{p^\prime} = \frac12\,\left(\mathbf{p^\prime}_1- \mathbf{p^\prime}_2\right)\,.
\end{equation}
The on-shell condition is given by $\mathbf{p}_+ \cdot \mathbf{p}_- = 0$.
The momenta are taken to be independent of \Nc, i.e., $p\sim \Lambda_{QCD}$.
Our analysis does not depend on a low-momentum expansion of the
potential, unlike in chiral or pionless EFTs.
The Hamiltonian $H$ is the nuclear Hamiltonian in the Hartree expansion, 
which in the large-\Nc limit can be written as \cite{Dashen:1994qi,Kaplan:1996rk}
\begin{equation}
H = \Nc \sum_{s,t,u} v_{stu} \left( \frac{S}{\Nc} \right)^s \left( \frac{I}{\Nc} \right)^t \left( \frac{G}{\Nc} \right)^u \ ,
\end{equation}
where the coefficients $v_{stu}$ are functions of the momenta
$\mathbf{p}_\pm$. The operators $S$, $I$, and $G$ are given by
\begin{equation}
S^i = q^\dagger \frac{\sigma^i}{2}q \,, 
\quad I^a = q^\dagger \frac{\tau^a}{2}q \,, 
\quad G^{ia} = q^\dagger \frac{\sigma^i\tau^a}{4}q \,,
\end{equation}
and when evaluated between single-nucleon states scale as \cite{Kaplan:1996rk}
\begin{equation}
\langle N^\prime \vert S^i \vert N \rangle \sim 1\,,
\quad \langle N^\prime \vert I^a \vert N \rangle \sim 1\,, 
\quad \langle N^\prime \vert G^{ia} \vert N \rangle \sim \Nc \,.
\end{equation}
In the large-\Nc formalism, it is consistent to interpret the potential as
originating from one-meson exchanges
\cite{Kaplan:1996rk,Banerjee:2001js,Cohen:2002im}.  In this picture, a factor
of $\pplus$ arises from relativistic corrections and is therefore suppressed by
the nucleon mass $m_N$.  Since $m_N$ scales as \Nc and we consider momenta
$\sim N_c^0$, each power of  $\pplus$ introduces a suppression by 1/\Nc. The
coefficients $v_{stu}$ are constructed such that the resulting Hamiltonian has
specific symmetry properties.  In the following, $H$ is rotationally invariant,
even under particle interchange, time-reversal odd, and we consider both
parity-odd and parity-even cases.  The transformation properties under
time reversal (T), parity (P), and particle interchange ($P_{12}$) of the
various building blocks are given in Tables~\ref{tab:spintrans} and
\ref{tab:isospintrans}. There, $[AB]^{ij}_2$ denotes the symmetric and traceless
rank-two tensor, constructed from the vector quantities $A^i$, $B^j$ as
\begin{equation}
[AB]^{ij}_2 \equiv A^i B^j + A^j B^i -\frac{2}{3}\delta^{ij}A\cdot B\,.
\end{equation}
For a review of how to construct the TCPC NN potential see \cite{Phillips:2013rsa}.
In the next Section we apply those methods to obtain the TV NN potentials. 
\begin{table}
\begin{center}
\begin{tabular}{ c  c  c  c  c  c  c  c  }
\hline
\hline
\hspace{2em}  &  \quad$\ \pplus \ $ \quad 
              &  \quad$\ \pminus \ $ \quad 
              & \quad$\vsig_1\cdot\vsig_2$ \quad
              & \quad$(\vsig_1 + \vsig_2)^z$ \quad
              & \quad$(\vsig_1 - \vsig_2)^z$ \quad
              & \quad$\vsig_1 \times \vsig_2$ \quad
              & \quad$[\sigma_1\sigma_2]^{ij}_2$ \quad \\
\hline
T \quad& - & + &  + & - & - & + & +  \\
P \quad& - & - &  + & + & + & + & +  \\
$P_{12}$ \quad& - & - &  + & + & - & - & + \\
\hline
\hline
\end{tabular}
\caption{Transformation properties of momenta and spin operators under time reversal (T), parity (P), and particle interchange ($P_{12}$).}
\label{tab:spintrans}
\end{center}
\end{table}
\begin{table}
\begin{center}
\begin{tabular}{ c  c  c  c  c  c }
\hline
\hline
\hspace{2em}  & \quad $\vtau_1\cdot\vtau_2$ \quad 
              & \quad $(\vtau_1 + \vtau_2)^z$ \quad 
              & \quad $(\vtau_1 - \vtau_2)^z$ \quad 
              & \quad $(\vtau_1 \times \vtau_2)^z$ \quad 
              & \quad $[\tau_1\tau_2]^{zz}_2$ \quad  \\
\hline
T\quad  & + & + & + & - & + \\
P\quad  & + & + & + & + & + \\
$P_{12}$\quad  &  + & + & - & - & + \\
\hline
\hline
\end{tabular}
\caption{Transformation properties of isospin operators under time reversal (T), parity (P), and particle interchange ($P_{12}$).}
\label{tab:isospintrans}
\end{center}
\end{table}


\section{The time-reversal-invariance-violating potentials}
\label{sec:potentials}

\subsection{The TVPV potential}
\label{sec:TVPV}

We first consider the TVPV potential.  By using the $1/N_c$-counting rules for
the momenta, spin, and isospin operators, as well as their transformation
properties under time reversal, parity, and particle interchange, we construct
the TVPV potential up to NNLO in the large-\Nc counting.
There is one operator structure at LO, $\mathcal{O}(\Nc)$,
\begin{equation}
V_{N_c}^{\slashed{T}\slashed{P}} = N_c\,U_{\slashed{T}\slashed{P}}^1(\pminus^2)\, \mathbf{p}_-\cdot\,(\vsig_1\,\tau_1^z - \vsig_2\,\tau_2^z) \, .
\label{tvpvLO}
\end{equation}
At NLO, $\mathcal{O}(N_c^0)$, five additional operators contribute,
\begin{equation}
\begin{split}
V_{N_c^0}^{\sT\sP} =& \, U_{\slashed{T}\slashed{P}}^2(\mathbf{p}_-^2)\, \mathbf{p}_-\cdot\,(\vsig_1 - \vsig_2) \\
& + U_{\slashed{T}\slashed{P}}^3(\mathbf{p}_-^2)\, \mathbf{p}_-\cdot\,(\vsig_1 - \vsig_2)\,\vtau_1\cdot\,\vtau_2 \\
& + U_{\slashed{T}\slashed{P}}^4(\mathbf{p}_-^2)\, \mathbf{p}_-\cdot\,(\vsig_1 - \vsig_2)\,[\tau_1\,\tau_2]_2^{zz} \\
& + U_{\slashed{T}\slashed{P}}^5(\mathbf{p}_-^2)\, \mathbf{p}_+\cdot\,(\vsig_1\times \vsig_2)\,[\tau_1\,\tau_2]_2^{zz}  \\
& + U_{\slashed{T}\slashed{P}}^6(\mathbf{p}_-^2)\, \mathbf{p}_+\cdot\,(\vsig_1 \times\, \vsig_2)\,\vtau_1\cdot\,\vtau_2 \, .
\end{split}
\label{tvpvNLO}
\end{equation}
The NNLO, $\mathcal{O}(N_c^{-1})$, operators are given by
\begin{equation}
\begin{split}
V_{N_c^{-1}}^{\slashed{T}\slashed{P}}
= N_c^{-1} &  \Big[  U_{\slashed{T}\slashed{P}}^7(\mathbf{p}_-^2)\, \mathbf{p}_-\cdot\,(\vsig_1\,\tau_2^z - \vsig_2\,\tau_1^z)   \\
& \,\,\,\, + U_{\slashed{T}\slashed{P}}^8(\mathbf{p}_-^2)\, \mathbf{p}_+^2\,\mathbf{p}_-\cdot\,(\vsig_1\,\tau_1^z - \vsig_2\,\tau_2^z) \\
& \,\,\,\, + U_{\slashed{T}\slashed{P}}^9(\mathbf{p}_-^2)\, \mathbf{p}_+\cdot\,(\vsig_1 + \vsig_2)\,(\vtau_1\times\,\vtau_2)^z  \\
& \,\,\,\, + U_{\slashed{T}\slashed{P}}^{10}(\mathbf{p}_-^2)\, \mathbf{p}_+\cdot\,(\vsig_1 \times\, \vsig_2)\,(\vtau_1 + \vtau_2)^z \\
&\,\,\,\,+ U_{\slashed{T}\slashed{P}}^{11}(\mathbf{p}_-^2)\,[(\mathbf{p}_+\times \mathbf{p}_-) \mathbf{p}_-]^{ij}_2
\,[\sigma_1 \sigma_2]^{ij}_2\,(\vtau_1 - \vtau_2)^z \\
&\,\,\,\, + U_{\slashed{T}\slashed{P}}^{12}(\mathbf{p}_-^2)\, [(\mathbf{p}_+\times \mathbf{p}_-) \mathbf{p}_+]^{ij}_2
\,[\sigma_1\, \sigma_2]^{ij}_2\,(\vtau_1\times\,\vtau_2)^z \Big]  \,.
\end{split}
\label{tvpvNNLO}
\end{equation} 
The $U_{\slashed{T}\slashed{P}}^i(\pminus^2)$ are arbitrary functions of
$\pminus\sim\Nc^0$ and do not change the large-\Nc scaling of the corresponding
operator structures. While corrections to the LO term in the potential are
suppressed by single powers of $1/\Nc$, for a given isospin sector the first
correction is suppressed by $1/\Nc^2$: the LO term in the potential is an isovector, 
while the $1/\Nc$-suppressed terms are purely
isoscalar and isotensor pieces. The NNLO contributions are again only of isovector
form.


\subsection{The TVPC potential}
\label{sec:TVPC}

The TVPC potential can be constructed analogously.
In this case, the LO contribution appears at $\mathcal{O}(\Nc^0)$, and is
therefore suppressed compared to the LO terms of the TVPV potential.
There are two LO operators, 
\begin{equation}
\begin{split}
V_{\Nc^0}^{\slashed{T}P} = & \, U_{\slashed{T}P}^1(\mathbf{p}_-^2)\, 
\mathbf{p}_-^i\,\mathbf{p}_+^j\,[\sigma_1\sigma_2]^{ij}_2\,\vec\tau_1\cdot\,\vec\tau_2  \\
&+ U_{\slashed{T}P}^2(\mathbf{p}_-^2)\, \mathbf{p}_-^i\,\mathbf{p}_+^j\,[\sigma_1\sigma_2]^{ij}_2\,[\tau_1\,\tau_2]^{zz}_2  \,.
\end{split}
\label{tvpcLO}
\end{equation}
The NLO, $\mathcal{O}(N_c^{-1})$, operators are given by
\begin{equation}
\begin{split}
V_{\Nc^{-1}}^{\slashed{T}P} =  \Nc^{-1} & \Big[ U_{\slashed{T}P}^3(\mathbf{p}_-^2)\, (\mathbf{p}_-\times\mathbf{p}_+)
\cdot\,(\vec\sigma_1\times\vec\sigma_2)\,(\vtau_1 - \vtau_2)^z \\
& \,\,\,\, + U_{\slashed{T}P}^4(\mathbf{p}_-^2)\, (\mathbf{p}_-\times\mathbf{p}_+)\cdot\,(\vec\sigma_1 - \vec\sigma_2)\,(\vec\tau_1\times\vec\tau_2)^z
\\
& \,\,\,\, +  U_{\slashed{T}P}^5(\mathbf{p}_-^2)\, \mathbf{p}_-^i\,\mathbf{p}_+^j\,[\sigma_1\sigma_2]^{ij}_2\,(\vtau_1 + \vtau_2)^z \Big] \,.
\end{split}
\label{tvpcNLO}
\end{equation}
For completeness, we also show the result for the NNLO, $\mathcal{O}(\Nc^{-2})$
operators, even though this order is not considered for the TVPV case, 
\begin{equation}
\begin{split}
V_{N_c^{-2}}^{\slashed{T}P}
=    N_c^{-2} & \Big[ U_{\slashed{T}P}^6(\mathbf{p}_-^2)\, \mathbf{p}_-^i\,\mathbf{p}_+^j\,[\sigma_1\sigma_2]^{ij}_2  \\
& \,\,\,\, + U_{\slashed{T}P}^7(\mathbf{p}_-^2)\, \mathbf{p}_+^2\,\mathbf{p}_-^i\,\mathbf{p}_+^j\,
[\sigma_1\sigma_2]^{ij}_2\,\vec\tau_1\cdot\,\vec\tau_2  \\
& \,\,\,\, + U_{\slashed{T}P}^8(\mathbf{p}_-^2)\, \mathbf{p}_+^2\,\mathbf{p}_-^i\,\mathbf{p}_+^j\,
[\sigma_1\sigma_2]^{ij}_2 \,[\tau_1\,\tau_2]^{zz}_2
\Big] \,.
\end{split}
\label{tvpcNNLO}
\end{equation}
The $U_{\slashed{T}P}^i(\pminus^2)$ are again arbitrary functions that do not
scale with $\Nc$. As in the TVPV case, for a given isospin sector the first
corrections are suppressed by $1/\Nc^2$, e.g., here the LO isoscalar and isotensor terms only get corrections at NNLO.


\section{Comparison with phenomenological TV potentials}
\label{sec:Compare}

In the following we compare our results with existing parameterizations of the
TV potentials and extract the large-\Nc scaling of the corresponding couplings.
If available at all, experimental constraints on the TV couplings
are very weak (see, e.g., Ref.~\cite{Herczeg:1995}), 
so we are unable to compare our results to data.  However, the
hierarchy of couplings established in our analysis should prove helpful in
identifying the most relevant couplings on which to focus in future TV studies.

\subsection{General parameterization}
\label{sec:Herczeg}

\subsubsection{TVPV potential}

A general parameterization of the TVPV and TVPC Hamiltonians  to first order in
$\pplus$ was given in Ref.~\cite{Herczeg:1966}.  We follow the notational
conventions of Ref.~\cite{Song:2011sw}, but adapt them to our definition of the
potential as a function of $\pminus$ and $\pplus$.  The resulting potential can
be written as 
\begin{align}
V_{\slashed{T}\slashed{P}} = & \, \Big[  \bg_1(\pminus^2) + \bg_2(\pminus^2) \, \vtau_1\cdot \vtau_2
+ \bg_3(\pminus^2) \, [\tau_1\,\tau_2]_2^{zz} \Big] \,  \pminus\cdot(\vsig_1-\vsig_2) 
\nonumber\\
& + \left( \bg_4 (\pminus^2) + \bg_5 (\pminus^2) \right) \, \pminus\cdot(\vsig_1 \tau_1^z -\vsig_2 \tau_2^z)
\nonumber\\
& + \left( \bg_4 (\pminus^2) - \bg_5 (\pminus^2) \right) \, \pminus\cdot(\vsig_1 \tau_2^z -\vsig_2 \tau_1^z)
\nonumber\\
& + \Big[ \bg_6 (\pminus^2) - \bg_{10} (\pminus^2)  
 + \big( \bg_7 (\pminus^2) - \bg_{11} (\pminus^2) \big) \, \vtau_1\cdot \vtau_2
\nonumber\\
& \quad + \big( \bg_8 (\pminus^2) - \bg_{12} (\pminus^2) \big)  \, [\tau_1\,\tau_2]_2^{zz}
 + \big( \bg_9 (\pminus^2) - \bg_{13} (\pminus^2) \big)  \, (\vtau_1 + \vtau_2)^z
\Big] \pplus \cdot (\vsig_1 \times\, \vsig_2) 
\nonumber\\
& + \bg_{14} (\pminus^2)  \, [(\mathbf{p}_+\times \mathbf{p}_-) 
\mathbf{p}_-]^{ij}_2 \,[\sigma_1 \sigma_2]^{ij}_2\,(\vtau_1 - \vtau_2)^z \nonumber\\
& + \left( \bg_{15} (\pminus^2) - \bg_{16} (\pminus^2) \right) \pplus \cdot (\vsig_1 + \vsig_2) \, (\vtau_1 \times \vtau_2)^z \, .
\label{VHerczeg}
\end{align}
The functions $\bg_i(\pminus^2)$ are related to Fourier transforms of the
functions $g_i(r)$ of Ref.~\cite{Song:2011sw}. Because $\pminus$ is independent
of $\Nc$, the Fourier transform does not alter the large-$\Nc$ scaling and the
relations derived below for the $\bg_i(\pminus^2)$ should also hold for
the corresponding $g_i(r)$. Comparison with
Eqs.~\eqref{tvpvLO}-\eqref{tvpvNNLO} shows that these structures are reproduced
in the large-\Nc analysis up to NNLO, with the exception of the term
proportional to $(\bg_6-\bg_{10})$, which is suppressed even further.
On the other hand, Eq.~\eqref{tvpvNNLO} contains an additional term,
proportional to $U_{\slashed{T}\slashed{P}}^{12}(\mathbf{p}_-^2)$, which is not
included in Eq.~\eqref{VHerczeg} because it is second order in $\pplus$.  The
following large-\Nc scaling relations for the couplings can be extracted:
\begin{align}
\bg_1 & \sim \Nc^0 \,, & \bg_2 & \sim \Nc^0 \, , & \bg_3 & \sim \Nc^0 \, ,\nonumber \\
\left(\bg_4+\bg_5\right) & \sim \Nc \, , & \left(\bg_4 - \bg_5\right) & \sim \Nc^{-1} \,, \nonumber\\
\left( \bg_6 - \bg_{10}\right)& \sim \Nc^{-2} \, , & \left( \bg_7 - \bg_{11}\right) & \sim \Nc^0 \, , & \left( \bg_8 - \bg_{12}\right) & \sim \Nc^0 \, , \nonumber\\
\left( \bg_9 - \bg_{13}\right) & \sim \Nc^{-1} \,, & \bg_{14} & \sim \Nc^{-1} \,, & \left( \bg_{15} - \bg_{16}\right) & \sim \Nc^{-1} \,.
\end{align}
In the large-\Nc limit, the order-$\Nc$ TVPV interactions proportional to $\bg_{4} +
\bg_{5}$ dominate. From the two relations containing $\bg_4$ and $\bg_5$ it follows that
these two couplings are equal up to corrections of relative order $1/\Nc^2$, i.e., up to
corrections expected to be of order 10\%:
\begin{equation} 
\bg_4 =
\bg_5 \left( 1 + \mathcal{O}(1/\Nc^2) \right) \,.  
\end{equation} 
Terms proportional to $\pplus$ are absent at LO and start to contribute at NLO, 
leading to the order-$\Nc^0$ scaling of $\left( \bg_7 - \bg_{11}\right)$ 
and $\left( \bg_8 - \bg_{12}\right)$, the same order as some of the terms in the static potential.

\subsubsection{TVPC potential}

The general parameterization of the TVPC  Hamiltonian up
to first order in the relative momentum contains 18 terms \cite{Herczeg:1966}.
Here we only show those that have a corresponding term in
Eqs.~\eqref{tvpcLO}-\eqref{tvpcNNLO}, following some of the notational
conventions of Ref.~\cite{Song:2011jh}.  The terms proportional to $\tilde{g}_1$ through $\tilde{g}_8$
vanish because of the on-shell condition $\pminus\cdot \pplus = 0$. The
potential can then be written as
\begin{align}
V_{\slashed{T}P} = & \, \Big[ \tilde{g}_9(\pminus^2) - \tilde{g}_{13}(\pminus^2) 
+ \big( \tilde{g}_{10}(\pminus^2) - \tilde{g}_{14}(\pminus^2) \big) \, \vtau_1\cdot\vtau_2
\nonumber\\
& \, \, + \big( \tilde{g}_{11}(\pminus^2) - \tilde{g}_{15}(\pminus^2) \big) \,  [\tau_1\,\tau_2]_2^{zz}
 + \big( \tilde{g}_{12}(\pminus^2) - \tilde{g}_{16}(\pminus^2) \big) \, (\vtau_1 + \vtau_2)^z
\Big] \, \pminus^i\,\pplus^j\,[\sigma_1\sigma_2]^{ij}_2
\nonumber\\
& + \tilde{g}_{17}(\pminus^2) \, (\pminus \times \pplus)  \cdot (\vsig_1 \times \vsig_2) (\vtau_1 - \vtau_2)^z
\nonumber\\
& + \tilde{g}_{18}(\pminus^2) \, (\pminus \times \pplus)  \cdot (\vsig_1 - \vsig_2) (\vtau_1 \times \vtau_2)^z.
\end{align}
Identifying the operators structures with those of
Eqs.~\eqref{tvpcLO}-\eqref{tvpcNNLO}, the following large-\Nc scalings for the
functions $\tilde{g}_i(\pminus^2)$ (we use the tilde to distinguish them from the TVPV
functions $\bg_i (\pminus^2)$) are extracted:
\begin{align}
\left(\tilde{g}_9 - \tilde{g}_{13}\right) & \sim \Nc^{-2}\,, 
& \left(\tilde{g}_{10} - \tilde{g}_{14}\right) & \sim \Nc^{0}\,, 
& \left(\tilde{g}_{11} - \tilde{g}_{15}\right)  & \sim \Nc^{0}\,, \nonumber\\
\left(\tilde{g}_{12} - \tilde{g}_{16}\right) & \sim \Nc^{-1}\,, & \tilde{g}_{17} & \sim \Nc^{-1}\,, & \tilde{g}_{18} & \sim \Nc^{-1}\,.
\end{align}
Contrary to what was observed in the TVPV case, in the TVPC potential 
terms proportional to $\pplus$ are already present at LO. This leads to 
a relative suppression of $1/N_c$, so that the dominant TVPC interactions proportional to $\tilde{g}_{10} -
\tilde{g}_{14}$ and $\tilde{g}_{11} - \tilde{g}_{15}$ are of order $\Nc^0$. Again, the next-order
terms are only suppressed by a single factor of 1/\Nc.  The terms proportional
to $U_{\slashed{T}P}^7(\pminus^2)$ and $U_{\slashed{T}P}^8(\pminus^2)$ in
Eq.~\eqref{tvpcNNLO} contain more than one power of $\pplus$ and thus were not
considered in Ref.~\cite{Herczeg:1966}.


\subsection{One-meson exchange potential}

\subsubsection{TVPV Potential}

The TVPV potential is commonly parameterized in terms of one-meson exchanges
with one TCPC and one TVPV meson-nucleon coupling \cite{Haxton:1983dq,Herczeg:1987,Gudkov:1992yc,Towner:1994qe}. 
Following Ref.~\cite{Liu:2004tq}, we consider $\pi$, $\eta$, $\rho$, and $\omega$
exchanges.  The Lagrangian describing the TCPC meson-nucleon interactions is
given by
\begin{align}
{\cal L}_\text{st}
=& \, g_{\pi}\bar{N} i\gamma_5\tau^a \pi^a N +g_{\eta }\bar{N}i\gamma_5\eta N
\nonumber\\
& - g_{\rho }\bar{N}\left(\gamma^\mu-i\frac{\xi_V}{2\Lambda}\sigma^{\mu\nu} q_\nu\right)\tau^a \rho^a_\mu N
- g_{\omega }\bar{N}\left(\gamma^\mu-i\frac{\xi_S}{2\Lambda}\sigma^{\mu\nu} q_\nu\right)\omega_\mu N \ , \label{eq:Lstrong}
\end{align}
where $q_\nu = p_\nu-p^\prime_\nu$,  while the TVPV Lagrangian reads
\begin{align}
{\cal L}_{\slashed{T}\slashed{P}}
=& \, \bar{N}\,\Big(\,\bar{g}_\pi^{(0)} \tau^a \pi^a+\bar{g}_\pi^{(1)}\pi^0
           +\bar{g}_\pi^{(2)}(3\tau^z\pi^0-\tau^a\pi^a)\,\Big)\,N
\nonumber\\
&+ \bar{N}\,\Big(\,\bar{g}^{(0)}_\eta\eta + \bar{g}^{(1)}_\eta \tau^z \eta\,\Big)\, N
\nonumber\\
&+ \bar{N}\,\Big(\, \bar{g}_\rho^{(0)}\tau^a \rho_\mu^a
                         +\bar{g}^{(1)}_\rho \rho^0_\mu
                         +\bar{g}_\rho^{(2)}(3\tau^z\rho_\mu^0-\tau^a\rho^a_\mu )\,\Big)\,
                         \frac{\sigma^{\mu\nu}q_\nu\gamma_5}{2\Lambda} N
\nonumber\\
&+ \bar{N}\,\Big(\,\bar{g}^{(0)}_\omega\omega_\mu
                         +\bar{g}^{(1)}_\omega \tau^z \omega_\mu\,\Big)\,
                         \frac{\sigma^{\mu\nu}q_\nu\gamma_5}{2\Lambda}\, N\,.
\label{tvpv-g}
\end{align}
In comparison to Ref.~\cite{Liu:2004tq} we have replaced $\chi_{V,S}/m_N\to
\xi_{V,S}/\Lambda$ in ${\cal L}_\text{st}$ and $1/m_N\rightarrow 1/\Lambda$
in ${\cal L}_{\slashed{T}\slashed{P}}$, where $\Lambda \sim 1\,\text{GeV}$ is
independent of \Nc.  
This prevents spurious factors of $m_N \sim \Nc$ from appearing in the
expression for the potentials; see Ref.~\cite{Phillips:2014kna} for an
analogous discussion for the TCPV case.  The TVPV potential derived from these
Lagrangians is given in Refs.~\cite{Liu:2004tq,Song:2011sw}.  Using our
conventions and transforming to momentum space it takes the form 
\begin{align}
V_{\slashed{T}\slashed{P}}^\text{meson} = &
\left[-\frac{\bar{g}^{(0)}_\eta g_\eta}{2 m_N}
               \,Y^{(\eta)}(\mathbf{p}_-^2)
               +\frac{\bar{g}^{(0)}_\omega g_\omega}{2\Lambda}
               \,Y^{(\omega)}(\mathbf{p}_-^2) \right]
               (\vec\sigma_1 - \vec\sigma_2)\cdot\pminus
\nonumber\\
&+\left[-\frac{\bar{g}^{(0)}_\pi g_\pi}{2 m_N}
              \,Y^{(\pi)}(\mathbf{p}_-^2)
              +\frac{\bar{g}^{(0)}_\rho g_\rho}{2\Lambda}
              \,Y^{(\rho)}(\mathbf{p}_-^2)\right]
              \vtau_1\cdot\vtau_2\,(\vec\sigma_1 - \vec\sigma_2)\cdot\pminus
\nonumber\\
&+\left[-\frac{\bar{g}^{(2)}_\pi g_\pi}{2 m_N}
              \,Y^{(\pi)}(\mathbf{p}_-^2)
              +\frac{\bar{g}^{(2)}_\rho g_\rho}{2\Lambda}
              \,Y^{(\rho)}(\mathbf{p}_-^2)\right]
              \frac{3}{2}[\tau_1\tau_2]_{2}^{zz}\,(\vec\sigma_1 - \vec\sigma_2)\cdot\pminus
\nonumber\\
&+\left[-\frac{\bar{g}^{(1)}_\pi g_\pi}{2 m_N}
               \,Y^{(\pi)}(\mathbf{p}_-^2)
              +\frac{\bar{g}^{(1)}_\omega g_\omega}{2\Lambda}
               \,Y^{(\omega)}(\mathbf{p}_-^2)\right]
               (\vec\sigma_1 \, \tau_1^z - \vec\sigma_2 \, \tau_2^z)\cdot\pminus
\nonumber\\
&+\left[ \frac{\bar{g}^{(1)}_\eta g_\eta}{2 m_N}
              \,Y^{(\eta)}(\mathbf{p}_-^2)
              -\frac{\bar{g}^{(1)}_\rho g_\rho}{2\Lambda}
              \,Y^{(\rho)}(\mathbf{p}_-^2)\right]
              (\vec\sigma_2\, \tau_1^z - \vec\sigma_1\, \tau_2^z)\cdot\pminus \,,
\label{tvpv-mesonpotential}
\end{align}
where $Y^{(a)}(\pminus^2) = \frac{1}{\pminus^2 + m_{a}^2}$.

Comparison of Eq.~\eqref{tvpv-mesonpotential} with
Eqs.~\eqref{tvpvLO}-\eqref{tvpvNNLO} shows that the meson-exchange potential
contains the LO term of Eq.~\eqref{tvpvLO}, as well as three of the five NLO
terms of Eq.~\eqref{tvpvNLO} and one NNLO term of Eq.~\eqref{tvpvNNLO}.  
Because the meson-exchange potential in the form
of Eq.~\eqref{tvpv-mesonpotential} is linear in the momenta and does not
include any relativistic corrections, it does not contain any of the operator
structures that are proportional to a single factor of $\pplus$, nor terms
that contain tensor structures of $\pminus$ and $\pplus$.

Now, using the known large-\Nc scalings of the strong couplings, it is possible to
determine the constraints that the large-\Nc analysis places on the TVPV
meson-nucleon couplings.  The \Nc scaling of the strong couplings is
\cite{Kaplan:1996rk,Banerjee:2001js,Phillips:2014kna}  
\begin{alignat}{2}
 g_{\pi} &\sim N_c^{3/2}\,,\quad & &g_{\eta} \sim N_c^{1/2}\,. \notag\\
 g_\omega &\sim N_c^{1/2}\,,\quad & &g_\omega\,\xi_S \sim N_c^{-1/2}\,, \notag \\
 g_\rho &\sim N_c^{-1/2}\,, \quad & &g_\rho\,\xi_V \sim N_c^{1/2}\, .
\end{alignat}
As stated above, the scale $\Lambda$ is independent of \Nc,
$\Lambda \sim \Nc^0$. The same holds for the momentum $\pminus$ and the meson masses $m_a$ ($a=\pi,\eta,\omega,\rho$), so we 
also have
\begin{equation}
Y^{(a)}(\pminus^2)\sim \Nc^0 \, .
\end{equation}
Requiring the coefficient functions $U_{\slashed{T}\slashed{P}}^i(\pminus^2)$
to be of order $N_c^0$ and not further suppressed, Eq.~\eqref{tvpv-mesonpotential}
allows to set constraints on 
the $N_c$ scalings of the TVPV meson-nucleon couplings.  
Because there are contributions of more than one TVPV coupling to a single operator structure in
Eq.~\eqref{tvpv-mesonpotential}, in
principle only upper limits can be extracted for their scaling. However, 
at large distances pions dominate compared to the heavier meson
exchanges. Therefore, pion couplings should saturate the upper limits
and we  obtain
\begin{align}
\bar{g}_\pi^{(0)}   & \sim \Nc^{-1/2}\,, & 
\bar{g}_\rho^{(0)}  & \lesssim \Nc^{1/2}\,, \notag \\
\bar{g}_\pi^{(1)}   & \sim \Nc^{1/2}\,, & 
\bar{g}_\omega^{(1)} & \lesssim \Nc^{1/2}\,, & &  \notag\\
\bar{g}_\pi^{(2)}   & \sim \Nc^{-1/2}\,, & 
\bar{g}_\rho^{(2)}  & \lesssim \Nc^{1/2}\,, \notag  \\
\bar{g}_\eta^{(0)}  & \lesssim \Nc^{1/2}\,, & 
\bar{g}_\omega^{(0)} & \lesssim \Nc^{-1/2}\,, \notag \\
\bar{g}_\eta^{(1)}  & \lesssim \Nc^{-1/2}\,, &
\bar{g}_\rho^{(1)} & \lesssim \Nc^{-1/2}\,.
\label{eq:TVPVmesonScale}
\end{align}
In the last two pairs of bounds obtained for $\bar{g}_\eta^{(0)}$, $\bar{g}_\omega^{(0)} $ 
and $\bar{g}_\eta^{(1)}$, $\bar{g}_\rho^{(1)} $ at least one of each pair of couplings 
must saturate the bound. 
In the pion sector a clear hierarchy between the various couplings is predicted. 
The isovector coupling
$\bar{g}_\pi^{(1)}$ dominates, while $\bar{g}_\pi^{(0)}$ and
$\bar{g}_\pi^{(2)}$ are both suppressed by a factor of $1/\Nc$, which 
agrees with the  $(\bar{g}_\pi^{(0)} - \bar{g}_\pi^{(2)}) \sim N_c^{-1/2}$ scaling found 
in the Skyrme model 
\cite{Riggs:1992jh}.

\subsubsection{TVPC Potential}

Constraints exist on the spin and parity of the exchanged bosons in the TVPC
potential, and these exclude, e.g., one-pion exchange \cite{Simonius:1975ve}.
Here we consider the potential of Ref.~\cite{Song:2011jh}, which includes
$\rho(770)$ and $h_1(1170)$ exchanges. These are the
lightest mesons that contribute to the TVPC potential. However, our analysis
can straightforwardly be extended if additional/different mesons are
considered.  The relevant interactions are \cite{Song:2011jh}
\begin{eqnarray}
{\cal L}_\text{st}&=&-g_{\rho}
  \bar{N}\left(\gamma^\mu-i\frac{\xi_V}{2\Lambda}\sigma^{\mu\nu} q_\nu\right)\tau^a \rho^a_\mu N
  -g_{h}{\bar N}\gamma^\mu\gamma_5 h_\mu N,
\nonumber\\
{\cal L}_{\slashed{T}P}&=&
- i \frac{\tilde{g}_\rho}{2\Lambda}
        {\bar N}\sigma^{\mu\nu}q_\nu(\vec{\tau} \times \vec\rho_{\mu})^z N
- \frac{\tilde{g}_h}{2\Lambda}
       {\bar N}\sigma^{\mu\nu}\gamma_5 q_\nu
       h_\mu N,
\end{eqnarray}
with the same replacements of $\chi_V/m_N \rightarrow \xi_V/\Lambda$ and
$1/m_N \rightarrow 1/\Lambda$  for the vector meson couplings as in the  TVPV
case. The potential in momentum space then reads (cf.~Ref.~\cite{Song:2011jh})
\begin{align}
\label{eq:vmesonTVPC}
V_{\slashed{T}P}^\text{meson} = & 
\frac{\tilde{g}_\rho g_\rho}{2m_N\Lambda}\, Y^{(\rho)}(\pminus^2) (\vtau_1\times\vtau_2)^z
                \;(\pminus \times \pplus)\cdot(\vec\sigma_1 - \vec\sigma_2)  \notag\\
& + \frac{\tilde{g}_h g_h}{2m_N\Lambda}\, Y^{(h)}(\pminus^2)
\mathbf{p}_-^i\,\mathbf{p}_+^j\,[\sigma_1\sigma_2]^{ij}_2 \ .
 \end{align}
To extract the large-\Nc scaling of the TVPC meson-nucleon couplings we take
the strong $hNN$ coupling to scale as \cite{Kaplan:1996rk,Banerjee:2001js} 
\begin{eqnarray}
g_{h} \sim N_c^{-1/2}\,.
\end{eqnarray}
Comparison with Eqs.~\eqref{tvpcLO}-\eqref{tvpcNNLO} shows that the $\rho$-meson exchange term corresponds to the NLO term proportional to
$U_{\slashed{T}P}^4(\mathbf{p}_-^2)$, while the $h_1$-meson term 
corresponds to the NNLO term
proportional to ${U}_{\slashed{T}P}^6(\mathbf{p}_-^2)$.  The TVPC
meson-nucleon couplings therefore scale as
\begin{equation}
\tilde{g}_\rho  \sim \Nc^{1/2} \,, \qquad \tilde{g}_h \sim \Nc^{-1/2} \,.
\end{equation}
The potential of Eq.~\eqref{eq:vmesonTVPC} does not contain any of the LO terms
in the large-\Nc counting. These are related to the exchange of additional
mesons.  For example, inclusion of the isovector $a_1$ meson results in a term
that matches the operator structure of the 
$U_{\slashed{T}P}^1(\pminus^2)$ term \cite{Song:2011jh}.  Given that the mass
of the $a_1(1260)$ is close to that of the $h_1(1170)$ meson, the large-\Nc
analysis suggests that $a_1$ exchange should not be neglected in
phenomenological applications.

\subsection{Effective field theory}

TVPV interactions have also been analyzed in effective field theory, see,
e.g.,
Refs.~\cite{deVries:2012ab,Bsaisou:2014oka,Mereghetti:2015rra,deVries:2015gea}
and references therein.  In a chiral EFT the interactions are parameterized in
terms of pion exchanges and nucleon-nucleon contact terms.  The 
LO potential is~\cite{deVries:2015gea}:
\begin{align}
V_{\slashed{T}\slashed{P}}^\text{EFT} 
= & - i \frac{\bar{C}_1}{2} \, (\vsig_1-\vsig_2) \cdot \pminus
- i \left( \frac{g_A [\bar{g}_\pi^{(0)} - \bar{g}_\pi^{(2)}]}{2 F_\pi}  \frac{1}{(\pminus^2+M_\pi^2)} 
+ \frac{\bar{C}_2}{2}\right) \, \vtau_1 \cdot  \vtau_2 \, (\vsig_1-\vsig_2) \cdot \pminus 
\nonumber \\
 & - i  \frac{g_A \bar{g}_\pi^{(1)}}{2 F_\pi}   \frac{1}{(\pminus^2+M_\pi^2)}  
(\vsig_1\tau_1^z - \vsig_2\tau_2^z)\cdot\pminus 
 \, .
\label{eq:VEFT}
\end{align}
Here $\bar{C}_{1,2}$ are NN contact terms, $F_\pi = 92.4 \ {\rm MeV}$ is the pion decay
constant, and $\bar{g}_\pi^{(0,1,2)}$ are the TVPV pion-nucleon couplings defined in Eq.~(\ref{tvpv-g}).
The term proportional to $\bar{g}_\pi^{(1)}$ in $V_{\slashed{T}\slashed{P}}^\text{EFT}$ 
reproduces the LO term in the large-\Nc
analysis. $V_{\slashed{T}\slashed{P}}^\text{EFT}$ also contains
two terms that are NLO in the 1/\Nc expansion.  This suggests
that, even though all three terms appear at the same order in chiral EFT,
the one-pion exchange contribution proportional to
$\bar{g}_\pi^{(1)}$ is dominant in a combined chiral and large-\Nc analysis. 
$\bar{g}_\pi^{(0,1,2)}$ are all assumed to be natural (i.e., of order 1) in the chiral EFT analysis, but
in fact $\bar{g}_\pi^{(0)}$ and $\bar{g}_\pi^{(2)}$ are suppressed compared to $\bar{g}_\pi^{(1)}$ by a factor of
$1/\Nc$.  Comparison with Eqs.~\eqref{tvpvLO}-\eqref{tvpvNNLO} also leads
to $\bar{C}_1 \sim N_c^0$,  $\bar{C}_2 \lesssim N_c^0$ for the NN contact 
terms. However, since
naturalness is difficult to define quantitatively and $1/\Nc = 1/3$ in the
physical world, it seems reasonable to retain all terms in Eq.~\eqref{eq:VEFT}
in phenomenological applications.

The fact that the LO chiral EFT potential in the TVPC case contains the leading term
in the $1/\Nc$ expansion is different from the TCPV case. There pion exchange
constitutes the sole LO contribution to the potential in the chiral 
counting, but the analysis of Ref.~\cite{Phillips:2014kna} shows it is actually suppressed by $\sin^2 \theta_W/N_c$
compared to other mechanisms.


\section{Conclusions}
\label{sec:conc}

We applied the $1/\Nc$ expansion to the TVPV and TVPC NN potentials. In
the TVPV case, the LO terms are of order $\Nc$, while the LO contributions in
the TVPC case are of order $\Nc^0$. In both cases first corrections are
suppressed by a single power of $1/\Nc$. However, to the order we considered,
the expansion in a given isospin sector is in $1/\Nc^2$, as it is in the TCPV and TCPC
cases \cite{Phillips:2014kna}.  In terms of a meson-exchange picture, the LO in
\Nc TVPV potential corresponds to $\pi$ and $\omega$ exchanges. Using the known
large-\Nc scaling of the strong meson-nucleon couplings, we derived bounds on
the scaling of the TVPV meson-nucleon couplings. In the pion sector, we find
that the isovector coupling $\bar{g}_\pi^{(1)}$ scales as $\Nc^{1/2}$, while
both isoscalar and isotensor couplings $\bar{g}_\pi^{(0)}$ and
$\bar{g}_\pi^{(2)}$ are smaller by a factor of $1/\Nc$. The NLO potential also contains
terms that are not reproduced in the meson-exchange picture. These  terms are
proportional to $\pplus$ and correspond to relativistic corrections.  In the
TVPC case, the commonly considered $\rho$ and $h_1$ exchanges only start to
contribute at NLO in the large-\Nc counting. The LO potential is generated by
the exchange of additional mesons, e.g., the $a_1$ meson. While these are
heavier than the $\rho$ and $h_1$ mesons, from the large-\Nc point of view all
meson masses scale as $N_c^0$ and the $a_1$ contribution should be considered.

Comparison with the TVPV potential $V_{\slashed{T}\slashed{P}}^\text{EFT}
$ derived at LO in chiral EFT shows that it reproduces the leading large-\Nc
operator, together with some subleading terms in
the large-\Nc expansion. In particular, the pion-exchange term
proportional to $\bar{g}_\pi^{(1)}$ contributes to the leading large-\Nc operator. This is
in contrast to the TCPV case, where the pion-exchange contribution, despite
being the LO term in the chiral power counting, only generates subleading terms
in the $1/\Nc$ expansion. 
The extracted large-\Nc scalings of the pion-nucleon couplings show that the TCPV pion-nucleon coupling $h^{(1)}_\pi$
is $1/\Nc$-suppressed relative to the TVPV pion coupling $\bar g^{(1)}_\pi$. 
This has the effect that the LO chiral TVPV single-pion exchange potential is enhanced compared to the LO chiral TCPV single-pion exchange.
It is interesting to note that, according to the
recent analysis of Ref.~\cite{Bowman:2014fca}, this
strong-interaction enhancement of the isovector TV pion exchange may increase the sensitivity
of experiments involving neutron scattering on nuclear targets to TV effects.

Given the difficulty of obtaining experimental constraints on the TV
couplings, future lattice QCD calculations, while themselves highly complex,
could contribute significantly to a better
understanding of CP-violating effects in nuclear systems. In particular,
calculations of the pion-nucleon couplings $\bar{g}_\pi^{(I)}$ ($I=0,1,2$) could check the hierarchy predicted 
by our large-\Nc analysis.

\begin{acknowledgments}
We thank V.~Gudkov, R.~P.~Springer, and W.~M. Snow for useful discussions.
This material is based upon work supported by the U.S. Department of Energy, Office of
Science, Office of Nuclear Physics, under Awards No.~DE-SC0010300 (MRS) and DE-FG02-93ER40756 (DRP).  
DS is supported by Thailand research fund TRF-RMUTI under contract No.~TRG5680079 and
Rajamangala University of Technology Isan.  DS, CS, and MRS thank the Institute
of Nuclear and Particle Physics at Ohio University for their hospitality.  CS
and MRS are also grateful to the Institute of Theoretical Physics II at the
Ruhr-Universit{\"a}t Bochum, Germany, for their hospitality.
\end{acknowledgments}


\end{document}